\documentclass[aps,prl,reprint]{revtex4-1}
\usepackage{graphicx}
\usepackage{xcolor}
\usepackage{mathtools}

\begin{document}
\title{Passive, broadband and low-frequency suppression of laser amplitude noise to the shot-noise limit using hollow-core fibre}
\author{Euan J. Allen\textsuperscript{*1, 2}}
\author{Giacomo Ferranti\textsuperscript{1}}
\author{Kristina R. Rusimova\textsuperscript{3}}
\author{Robert J. A. Francis-Jones\textsuperscript{3, 4}}
\author{Maria Azini\textsuperscript{3}}
\author{Dylan H. Mahler\textsuperscript{1, 5}}
\author{Timothy C. Ralph\textsuperscript{6}}
\author{Peter J. Mosley\textsuperscript{3}}
\author{Jonathan C. F. Matthews\textsuperscript{1}}
\affiliation{\textsuperscript{1}Quantum Engineering Technologies Labs, H. H. Wills Physics Laboratory and Department of Electrical \& Electronic Engineering, University of Bristol, BS8 1FD, United Kingdom}
\affiliation{\textsuperscript{2}Quantum Engineering Centre for Doctoral Training, Nanoscience and Quantum Information Centre, University of Bristol, BS8 1FD, United Kingdom}
\affiliation{\textsuperscript{3}Centre for Photonics and Photonic Materials, Department of Physics, University of Bath, Bath, BA2 7AY, United Kingdom}
\affiliation{\textsuperscript{4}Now at Clarendon Laboratory, University of Oxford, Parks Road, Oxford, OX1 3PU, United Kingdom}
\affiliation{\textsuperscript{5}Now at Xanadu, 372 Richmond St W, Toronto, ON M5V 2L7, Canada}
\affiliation{\textsuperscript{6}Centre for Quantum Computation and Communication Technology, School of Mathematics and Physics, University of Queensland, Brisbane, Queensland 4072, Australia} 
\affiliation{{*}euan.allen@bristol.ac.uk}

\begin{abstract}
We use hollow-core fibre to preserve the spectrum and temporal profile of picosecond laser pulses in a collinear balanced detection (CBD) scheme to suppress 2.6~dB of amplitude noise at MHz noise frequencies, to within 0.01~dB of the shot-noise limit. We provide an enhanced version of the CBD scheme that concatenates circuits to suppress over multiple frequencies and over broad frequency ranges --- we perform a first demonstration that reduces total excess amplitude noise, between 2~-~6~MHz, by 85\%. 
These demonstrations enable
passive, broad-band, all-guided fibre laser technology operating at the shot-noise limit.
\end{abstract}

\maketitle





Doped fibre lasers are attractive due to their low cost, compact size and robust stable operation. However, they exhibit classical intensity noise well above the optical shot-noise limit due to stimulated spontaneous emission ~\cite{budunoglu2009fiber,newbury2007low,paschotta2004noise,yue2013intensity}. To suppress this extra noise, 
existing noise suppression techniques could in principle 
enable use of fibre lasers for ultra-sensitive applications. However, direct changes to the laser itself --- such as intracavity spectral filtering~\cite{sanders1992reduction} and optical feedback into the laser cavity~\cite{solgaard1993optical,lang1980external} --- or external methods --- such as feedback or feedforward circuits coupled to an optical modulator~\cite{alnis2008subherts,robertson1986intensity} or external cavity filtering~\cite{hamilton1989introduction} --- all require fast and low-noise electronics. The speed of such electronics limits the bandwidth that can be filtered, while electronic noise can be transferred back into the optical beam. 
In practice, the ability of these schemes to reach the shot-noise limit can be hindered by fundamental limitations of two beams output from a beamsplitter being uncorrelated on the quantum level~\cite{bachor2004guide}.
They can also suffer from technical issues such as non-zero time response of any feedback mechanism~\cite{taubman1995intensity}, beam-geometry and pointing issues~\cite{robertson1986intensity}, and the tradeoff when tailoring the speed or gain of the feedback mechanism to reduce noise across a large bandwidth~\cite{hamilton1989introduction,alnis2008subherts,robertson1986intensity,taubman1995intensity}.   
Removing common mode noise using balanced detection~\cite{sonnenfroh2001application} is passive, however optical intensity must be equally split across the two detectors in order to remove all classical noise; this 
is
impossible 
for optical transmission measurements or imaging 
whenever the
sample 
introduces unknown loss prior to detection. 

\begin{figure}[tp!]
\centering
\includegraphics[width=\linewidth]{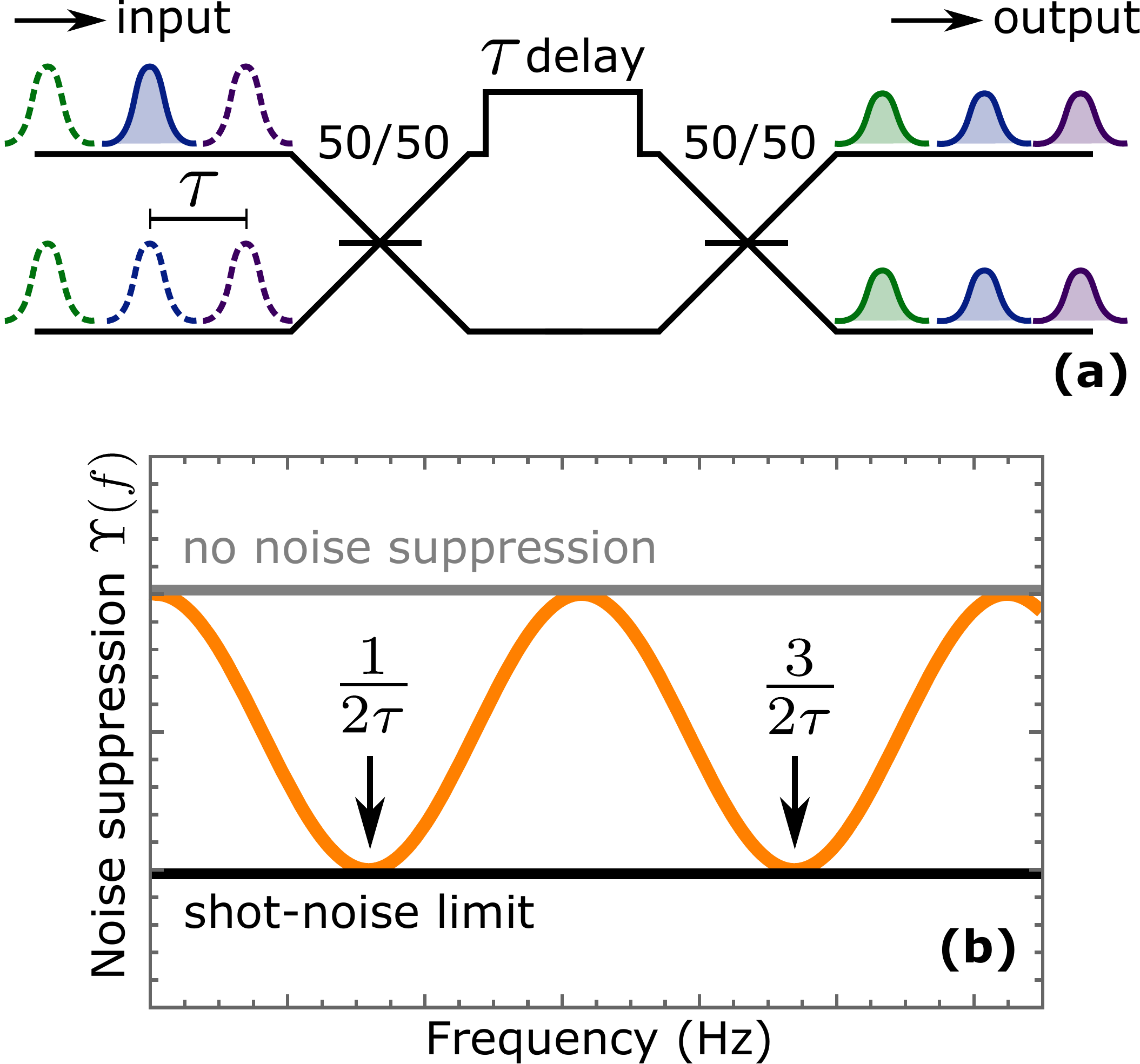}
\caption{\textbf{Collinear balanced detection.} (a) Pulsed light is input on the upper mode and split at a balanced beamsplitter; the top path is then delayed by time $\tau$ with respect to the bottom path, but such that the pulses from each path within the AMZI do not overlap in time at the second beamsplitter. This results in a pulse train at the output of the interferometer that has double the repetition rate, half the average power and a quarter of the peak power of the original pulse train. (b) Theoretical plot (Eq.~\ref{eq:noisesuppression}) of classical white noise suppressed by CBD over a range of noise frequencies given a single delay of $\tau$.  
CBD causes complete suppression odd multiples of the frequency $f = 1/(2\tau)$ at both outputs. Colour online. \label{figure1}}
\end{figure}
\begin{figure*}[htp!]
\centering
\includegraphics[width=\textwidth]{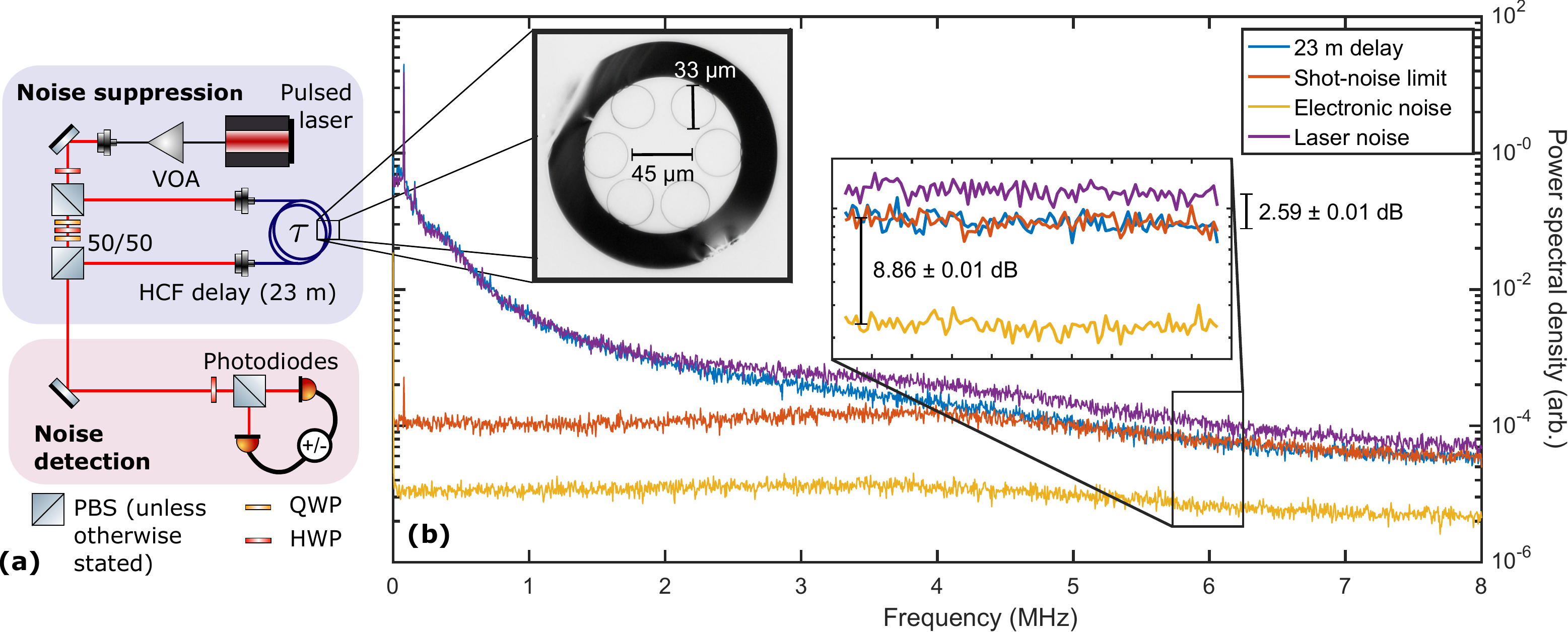}
\caption{
\textbf{Noise suppression using hollow core free-boundary fibre.} (a) The optical setup used to construct the asymmetric Mach-Zender interferometer used for noise suppression and the balanced detection technique to measure optical noise. Inset shows the structure of the free-boundary fibre. (b) Measurements taken with the setup, removing all classical noise at the expected frequency of $\sim 6.5$~MHz reaching to within 0.011~$\pm$~0.008~dB of the shot-noise limit (averaging the signal in a 0.5~MHz bandwidth centred on $6.5$~MHz). Displayed are the noise in the laser without filtering (purple), with filtering (blue), the equivalent power shot-noise limit (red), and electronic noise floor (yellow). The optical nose data here is included in Fig.~\ref{figure5}~b), plotted as a noise suppression ration (eq.~\ref{eq:noisesuppression}). The common mode rejection ratio for detection in this experiment is $>34$~dB and all measurements were performed with a total of 330~$\mu$W incident on the balanced detector. Colour online.\label{figure3}}
\end{figure*}

Collinear balanced detection (CBD) is a passive method that requires no knowledge of beam intensity and no intensity subtraction~\cite{nose2012sensitivity}. It comprises 
an optical delay in a static asymmetric Mach-Zehnder interferometer (AMZI) to provide an output beam with shot-noise limited intensity fluctuation at one frequency. This has been used, with a free space delay of 8.6~m, to increase the quality of stimulated Raman microscopy measured at 17.5~MHz~\cite{nose2012sensitivity} and to suppress classical noise from a noisy fibre laser used to generate amplitude-squeezed light with intensity fluctuations below the optical shot-noise limit at $20-24$~MHz~\cite{sawai2013photon}. However, two criticisms of CBD are that existing demonstrations have used free-space delays which
require the stability of optical isolation tables, and that existing CBD demonstrations suppress noise at only one frequency~\cite{freudiger2014stimulated}. Here we address both points. 

The original CBD concept is illustrated in Fig.~\ref{figure1}~(a). A train of laser pulses are launched into one input of an AMZI, where one path inside imparts a relative time delay of $\tau = L n/c$ ($L$ is the length of delay medium of refractive index $n$) such that the pulses in the delayed and non-delayed paths do not overlap in time when their paths re-combine at the second beamsplitter. Note that because the pulses do not optically
interfere at the second beamsplitter, the AMZI requires no phase stabilisation. 
Output from the CBD circuit, for a given $\tau$,
are two beams that both have a classical noise suppression factor $\Upsilon$, which follow a sinusoidal function (Fig.~\ref{figure1}~(b)) that reaches the shot-noise limit at odd multiples of $f=1/{2\tau}$~\cite{nose2012sensitivity},
\begin{equation}
\Upsilon (f) = \frac{1}{2} \left( 1 + \cos[\pi f \tau] \right),
\label{eq:noisesuppression}
\end{equation}
where $\Upsilon (f) = 1$ or $\Upsilon (f) = 0$ indicates that either zero or total classical noise suppression has occurred respectively (the latter becoming shot-noise limited). To remove noise at a frequency $f$, a delay of length $L = c/(2 n f)$ should be implemented.

The inverse relation with the suppression frequency ($\tau = 1/2f$) means the delay required for lower frequency noise suppression becomes increasingly challenging to build and maintain in free-space. For example, suppressing noise at 5~MHz requires 30~m of free-space delay. Constructing a waveguided delay long enough to suppress noise at low frequencies therefore makes CBD more practical and would allow integration of passive all-guided noise suppression with fibre lasers.
Solid-core optical fibre, such as 1550~nm SMF28, has the potential to filter noise at sub-$\text{kHz}$ frequencies thanks to the availability of lengths of many kilometres --- we illustrate this in the appendix with results for a 4.3~km delay that suppresses amplitude noise at 25~kHz. 
However, self-phase modulation and dispersion occur when laser pulses traverse solid-core fibre, which for the CBD scheme results in half of the output pulses being spectrally broadened and temporally elongated --- this is undesirable when the filtered light is to be used for applications sensitive to the spectral and temporal profile of laser pulses, such as nonlinear spectroscopy or microscopy. Furthermore, Raman scattering in solid core fibre will cause extra noise in the output optical beam of the CBD circuit~\cite{schmitt1998photon}, which will negate the noise suppression goal.

\begin{figure}[hbtp]
\centering
\includegraphics[width=\linewidth]{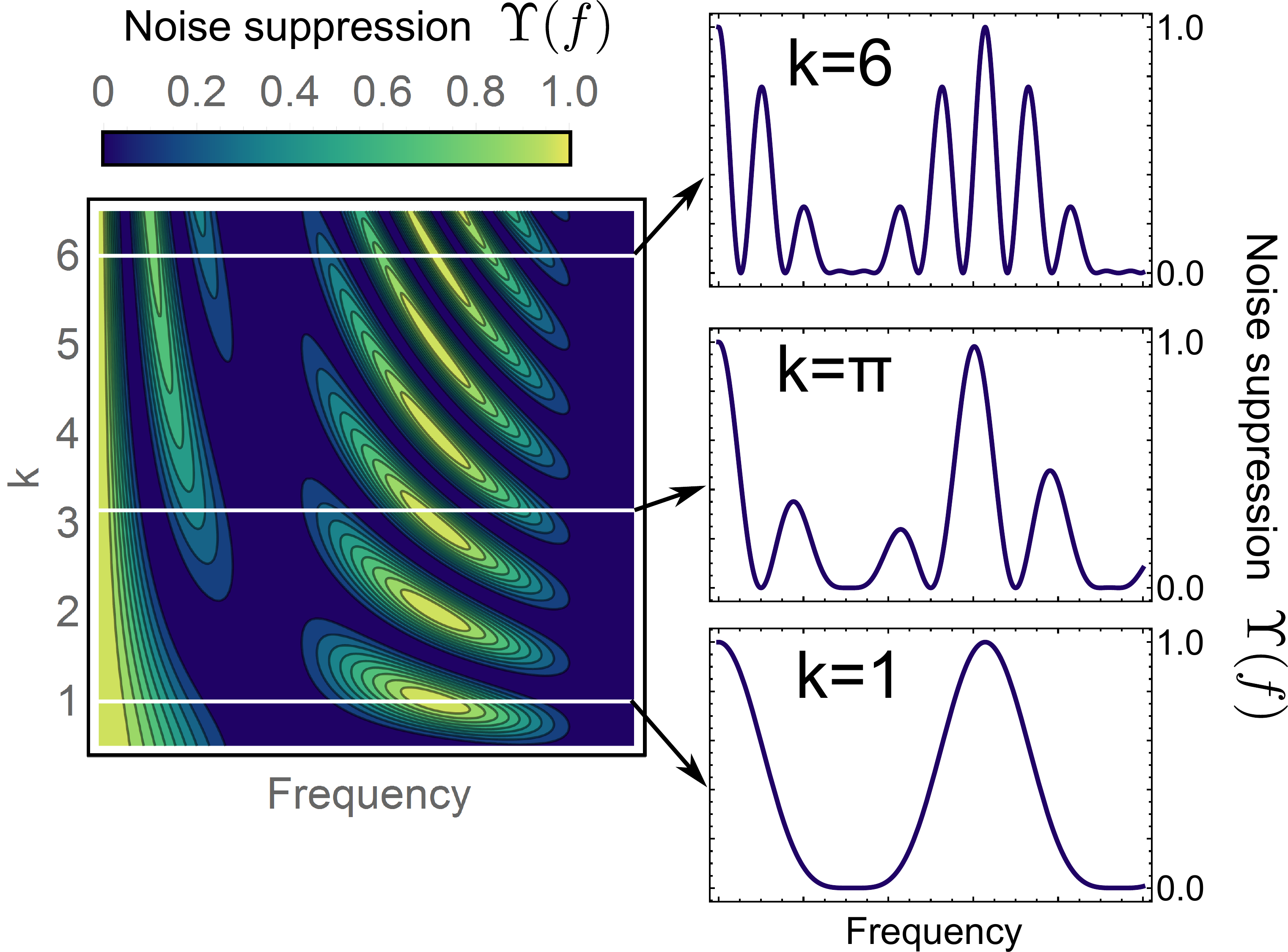}
\caption{
\textbf{Theoretical prediction for double-CBD scheme}. All plot show the amount of classical noise suppression from two noise suppressing interferometers with delays $\tau_1$ and $\tau_2 = k \tau_1$ respectively. Shown is both a contour plot of the predicted noise suppression for arbitrary $k = \tau_2/\tau_1$, with inset examples for fixed $k= \{6,\pi,1\}$. Colour online.
\label{figure4}}
\end{figure}

\begin{figure*}[hbtp]
\centering
\includegraphics[width=\textwidth]{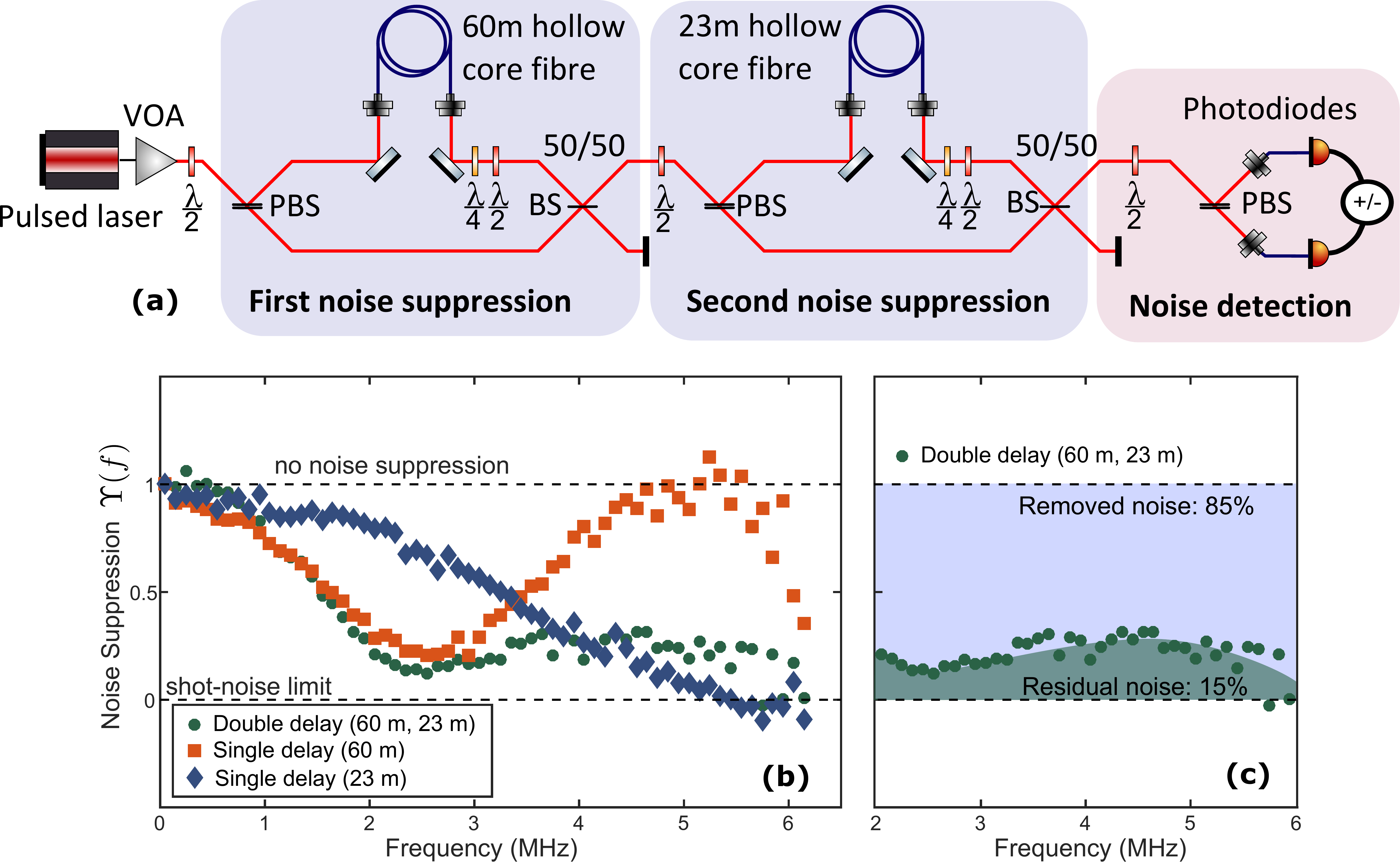}
\caption{
\textbf{Double-CDB circuit.} (a) Experimental setup of two sequential CBD circuits. 
Each interferometer is constructed as per Fig.~\ref{figure3}, but with now two lengths of 60~m and 23~m used as the delays. (b) Experimental results of the performance the double-CBD circuit, for the single delay of 23~m (blue), single delay of 60~m (orange) and both delays (green) in operation in the setup. We note the 60~m single delay suppresses noise at a frequency that is almost one order of magnitude lower than previous demonstrations~\cite{nose2012sensitivity, sawai2013photon}. The 23~m single delay data is re-plotted from Fig.~\ref{figure3}~b). (c) Repeat plot of experimental results for double-CDB circuit from (b), plotted between 2 - 6~MHz illustrating the 85\% total noise reduction achieve across this bandwidth. Colour online.\label{figure5}}
\end{figure*}

To avoid the nonlinear effects of conventional fibre, we use hollow core fibre (HCF) that guides light predominantly in an air-filled core. Here we used recently-developed free-boundary (or anti-resonant) HCF which has a simple cladding structure comprising of a single ring of silica capillaries fused to a fibre jacket (inset of Fig.~\ref{figure3}). The fibre guides light along the core via resonant trapping through the capillary structures~\cite{litchinitser2002antiresonant}. Hence, low-loss guidance along the fibre can be achieved across broad wavelength ranges~\cite{bird2017attenuation,yu2012fiber}. The fundamental mode propagates in a large core with almost no overlap with the glass to minimise nonlinearity and dispersion (see appendix), while careful control of the capillary dimensions during fabrication enables higher-order modes to be suppressed yet bend loss in the fundamental mode remain manageable~\cite{belardi2014hollow}. This simple structure enables fabrication of hundreds of metres of fibre with consistent properties and high damage threshold~\cite{jaworski2013picosecond}. 

Hollow core fibre fabrication was by the stack-and-draw process, forming a fibre of six capillaries with an average diameter of 33 $\mu$m surrounding a 45 $\mu$m core. The fibre had an outer diameter of 160 $\mu$m and the thickness of the capillary walls are estimated to be 520 nm. The attenuation of the fundamental mode in the fibre at a wavelength of 1550 nm was found by a cutback measurement on 34~m of fibre to be approximately 0.07 dB/m. By comparison with lengths of conventional single-mode fibre (SMF28), dispersion of the HCF was estimated to be 1.5~ps/nm/km, 9~\% of that for conventional single mode fibre (see appendix). To minimise bend loss within the HCF, the fibre was laid out on a piece of card in a circle of approximately 60~cm in diameter. 

We first constructed the CBD circuit in Fig.~\ref{figure3} (a) with a 23~m HCF delay. A half-waveplate and polarising beamsplitter replaces each balanced beamsplitter to fine tune the splitting ratio to compensate for any unbalanced loss in the delay path. Detection of the amplitude noise and shot-noise of the laser light was performed using self-homodyne~\cite{schmitt1998photon,ulrich2007integrated}, where the incident light is split on a balanced beamsplitter and detected by two identical photodiodes (in this paper: Thorlabs FGA01FC with custom amplification circuits). The signals from each photodiode are first filtered via DC (Mini-Circuits BLK-89-S+) and low-pass (Mini-Circuits BLP-21.4+) electronic filters and are then sent to individual channels on a Keysight InfiniiVision MSOX3104A oscilloscope (operated at 2.5 GSa/s). The low-pass filters remove high-frequency noise from the detectors allowing for higher resolution of the optical noise at the frequencies of interest. The DC filter allows for comparisons of noise at different incident powers to be made without concern for the offset change on the oscilloscope signal. All data captures are taken with a time resolution of 20~$\mu$s and with maximum voltage resolution possible for the amplitude of signal (typically 5 - 10~mV/division).  The signals are logged on a desktop PC, digitally added or subtracted from one another, and then Fourier transformed to the frequency domain to allow comparison of particular Fourier components of the noise. Subtraction of the photocurrents provides a shot-noise reference for the incident power and addition of the photocurrents provides the noise present in the incoming laser light. 

Light from the laser is first attenuated by a digital VOA (Oz Optics DA-100) to attenuate the optical power to work within the range of the detection system, before being launched into free space. The light is split into each path of the AMZI using a half-waveplate and polarising beamsplitter. Waveplates in the non-delayed path match the polarisation of the two paths. The pulse trains are recombined on a second (non-polarising) beamsplitter and the noise characteristics measured. The output of the SMF28 fibre from the pulsed fibre laser is coupled into free space via a 3-axis stage system (Elliot Scientific) with an aspheric lens system ($f=3.1$~mm, Thorlabs C330TMD-C). Coupling into the HCF used an identical system but with a $f=11.0$~mm (Thorlabs C220TMD-C) lens to allow for mode matching into the large core.

The laser used is an erbium-doped silica fibre laser that is passively mode-locked with a fibre coupled output and generates 2~ps pulses centred at wavelength 1550~nm with a repetition rate of 50~MHz (Pritel FFL). Without operation of the CBD circuit, our laser exhibits super-Poissonian amplitude noise present at all frequencies within the detector bandwidth (0 - 10~MHz). This is shown in Fig.~\ref{figure3} (b) --- the raw laser light (purple) contains noise contributions that are larger than the shot-noise limit (red). The laser's raw noise was measured by blocking one of the paths of the interferometer and then reducing the attenuation of the variable optical attenuator (VOA) to keep the power incident on the detectors the same in both suppressed and non-suppressed configurations. The total loss in the delay path is 3.6{~$\pm$~}0.2~dB, of which approximately 1~dB is from coupling in and out of the HCF and 1.6~dB from propagation loss. With the HCF-CBD circuit in operation, amplitude noise is suppressed as shown in Fig.~\ref{figure3}~(b) (blue), reaching to within 0.011~$\pm$~0.008~dB of the shot-noise limit at 6.5~MHz, as predicted for the length of fibre used. 

Next, we address the weakness that the original CBD scheme only filters at 
odd multiples of $f=1/{2\tau}$~\cite{freudiger2014stimulated}. We use the stability of fibre-delay to concatenate CBD circuits in series. As any individual CBD circuit does not add noise at any particular frequency, we find that concatenating two CBD circuits, each with delays $\tau_1$ and $\tau_2$, noise suppression follows 
\begin{equation}
\Upsilon (f) = \frac{1}{4} (1 +  \cos[\pi f \tau_1])(1 +  \cos[\pi f \tau_2]).
\label{UpsilonTwo}
\end{equation} 
The theoretical noise suppression across a fixed band of frequencies is shown in Fig.~\ref{figure4}. 

To explore this for instances of $\tau_1$ and $\tau_2$, we built the setup in Fig.~\ref{figure5} (a). Fig.~\ref{figure5} (b) displays the experimentally measured values of $\Upsilon (f)$, using a half-wave plate and polarising beamsplitter as tuneable beamsplitters to switch between only one of each CBD circuit suppressing noise and both circuits suppressing noise simultaneously. The total noise suppression across the 2 - 6~MHz bandwidth for the double-CBD scheme is 85\% of the original laser noise, which is shown in Fig.~\ref{figure5} (c). For an equivalent bandwidth, the 23~m and 60~m single-CDB implementations each remove less noise as expected: 55\% and 41\% respectively.

These demonstrations show free-boundary HCF can be used to implement CBD to suppress amplitude noise from a fibre laser at MHz frequencies. The cost of each CBD circuit is loss of half of the average power and three quarters of the peak power --- the same as passive temporal multiplexing used to increase laser repetition rate~\cite{Broome:11}. This suggests use where an overhead in power exists and the need for low amplitude noise takes priority, for example in ultra sensitive imaging~\cite{taylor2016quantum,freudiger2014stimulated,gambetta2010fiber,arroyo2016non} and squeezing experiments~\cite{an-physscript-91-053001}. While solid core fibre introduces the practicality of a waveguided optical delay, the detrimental nonlinear effects that it has
on laser pulses prohibit its use in CBD. Instead, the low optical nonlinearity of air-guiding safeguards spectral and temporal properties of output laser pulses, whilst providing the practicality and stability of a waveguide to further enables concatenation of sequential CBD circuits to suppress amplitude noise at multiple frequencies.
Further improvement of the HCF will allow longer delay to be used to suppress at sub-MHz frequencies --- iterative design of the fibre structure could reduce propagation loss and lower nonlinearity and dispersion, or create anomalous regimes where soliton formation is possible~\cite{knight2003photonic}. 
Efficient interface between solid core fibre and HCF will be needed for fully-integrated noise suppression, and this has been shown 
to be possible with as little as 0.3~dB splice loss~\cite{arc2006thapa}. Generalisation of CBD itself to continuous-wave light and temporal overlap of the pulses within the interferometer will further widen application and is the subject of future study.

\bibliographystyle{apsrev4-1} 

\section{Acknowledgements}
We are grateful to for assistance on detector electronics Yu Shiozawa. 
This work was supported by EPSRC programme grant EP/L024020/1, EPSRC UK Quantum Technology Hub in Quantum Enhanced Imaging (EP/M01326X/1), US Army Research Office (ARO) grant no. W911NF-14-1-0133, the EPSRC UK Quantum Technology Hub in Networked Quantum Information Technologies (EP/M013243/1), Innovate UK project FEMTO-AAD (102671), the Australian Research Council (ARC) under the Centre of Excellence for Quantum Computation and Communication Technology (CE170100012), and the Centre for Nanoscience and Quantum Information (NSQI).
EA was supported by the Bristol Quantum Engineering Centre for Doctoral Training, EPSRC Grant No. EP/L015730/1. JCFM acknowledges support from an EPSRC Early Careers Fellowship (EP/M024385/1) and an ERC starting grant ERC-2018-STG 803665. 

\null\clearpage

\onecolumngrid

\section{Appendix}
\subsection{Solid core all-fibre integration}
Here we show and discuss the capabilities of an all-guided fibre implementation of the collinear balanced detection scheme. The benefits of an all-fibre experimental implementation over previous free space strategies~\cite{nose2012sensitivity,sawai2013photon} is that propagation in fibre allows long delays to be implemented (and hence low frequency noise to be suppressed) with relative ease, as no alignment in necessary. The schemes inherent lack of requirement of phase stability also means that an all-fibre solution offers a simple `plug-and-play' option to suppressing noise. The scheme requires no alignment, a handful of components, and is robust against environmental challenges such as changes in temperature or vibrations. This type of interferometer also allows simple integration with current fibre laser systems. 

\begin{figure}[hbtp]
\centering
\includegraphics[width=\textwidth]{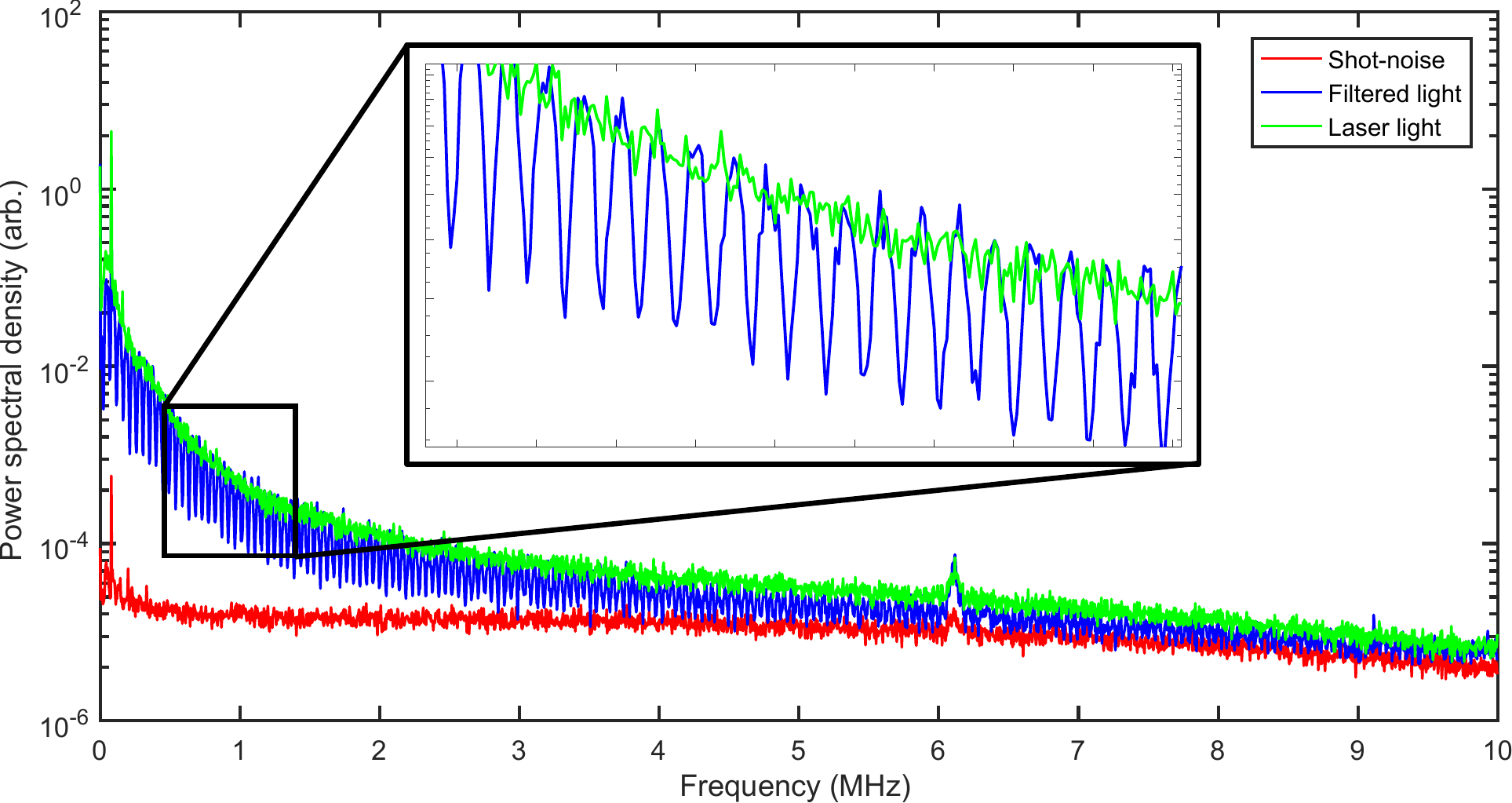}
\caption{Noise suppression produced by an all-guided fibre implementation of the collinear balanced detection scheme. The delay was implemented using 4.3~km of single-mode fibre which allows for noise suppression to occur at very low frequencies (approximately 25~kHz).\label{figure2}}
\end{figure}

Figure~\ref{figure2} demonstrates the noise suppression achieved when a delay of 4.3~km is implemented in single-mode solid core (SMF28) fibre. The 50/50 beamsplitters were implemented using fibre beamsplitters (Thorlabs TW1550R5A2) and an in-line polarisation controller (FiberPro PC1100 series) was used to correct the polarisation mismatch between pulses that travelled along the delay and those that did not. The laser was a Pritel Femtosecond Fiber Laser (Pritel FFL); an Er-doped silica fibre laser that is passively mode-locked with a fibre coupled output. The laser produces 2~ps pulses centred on a wavelength of 1550~nm at a repetition rate of 50~MHz. A digital variable optical attenuator (VOA, Oz Optics DA-100) was used at the output of the laser system to attenuate the optical power to work within the range of the detection system. A polarisation control and polarising beam splitter (PBS) combination was used in detection to ensure perfect balancing of power onto the two photodiodes. This ensures best extinction of classical laser noise in the subtraction measurement, producing an accurate shot-noise calibration.

\begin{figure}[hbtp]
\centering
\includegraphics[width=\textwidth]{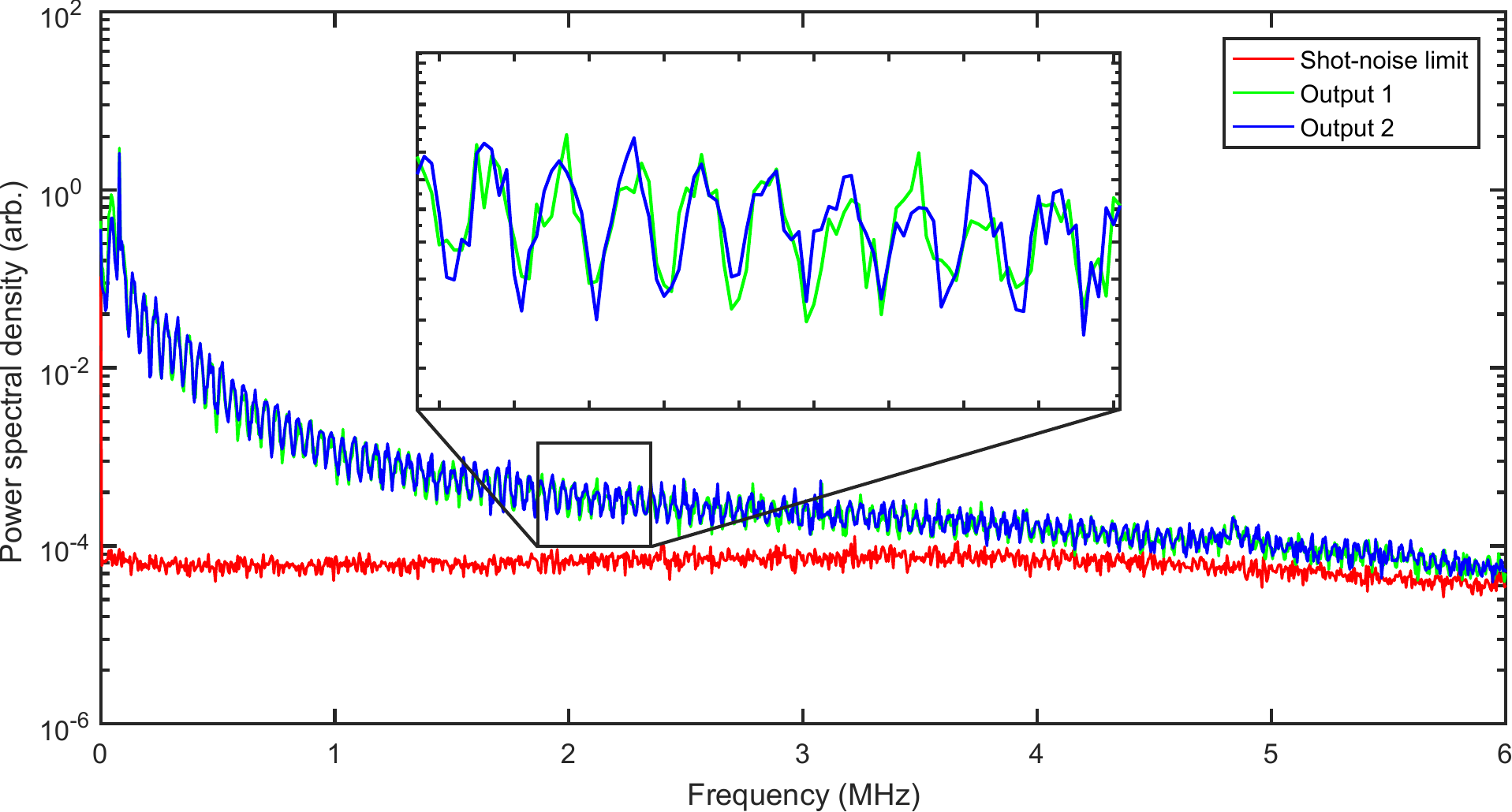}
\caption{Noise characteristics at both interferometer outputs as a function of frequency. We see that both ports are reduced in noise simultaneously.\label{fig:both}}
\end{figure}

Detection of the amplitude noise and shot-noise of the laser light was performed using the same technique as used in the results in the main paper. The light produced by the Pritel laser system has classical noise present at all frequencies within the detector bandwidth (0 - 10~MHz). This can be observed in Figure~\ref{figure2} that shows that the laser light photocurrent addition (green) contains larger spectral density components for all frequencies than the shot-noise (red, photocurrent subtraction). The noise suppressed light in Figure~\ref{figure2} (blue) demonstrates that a long delay can filter noise at low frequencies (25~kHz for 4.3~km) and also that the noise suppression follows the cosine relationship predicted. We note that the scheme does not reach the shot-noise limit at all frequencies due to the increase loss in the delayed arm, which was not compensated for in this case. Figure~\ref{fig:both} shows that the noise suppression occurs in both output arms simultaneously.

\subsection{Hollow core fibre dispersion and non-linearity}
Here we show measurements of the dispersion and nonlinearity of the hollow core fibre used in the primary experiment of the paper. The dispersion is estimated using a Femtochrome FR103-PD autocorrelator and comparing the the pulse broadening with the hollow core fibre with that through conventional single mode fibre. The nonlinearity is estimated by looking for nonlinear broadening in the optical spectrum after propagation along the fibre delay. The spectra were measured using an Anritsu MS9740A optical spectrum analyser.

\begin{figure}[hbtp]
\centering
\includegraphics[width=0.9\linewidth]{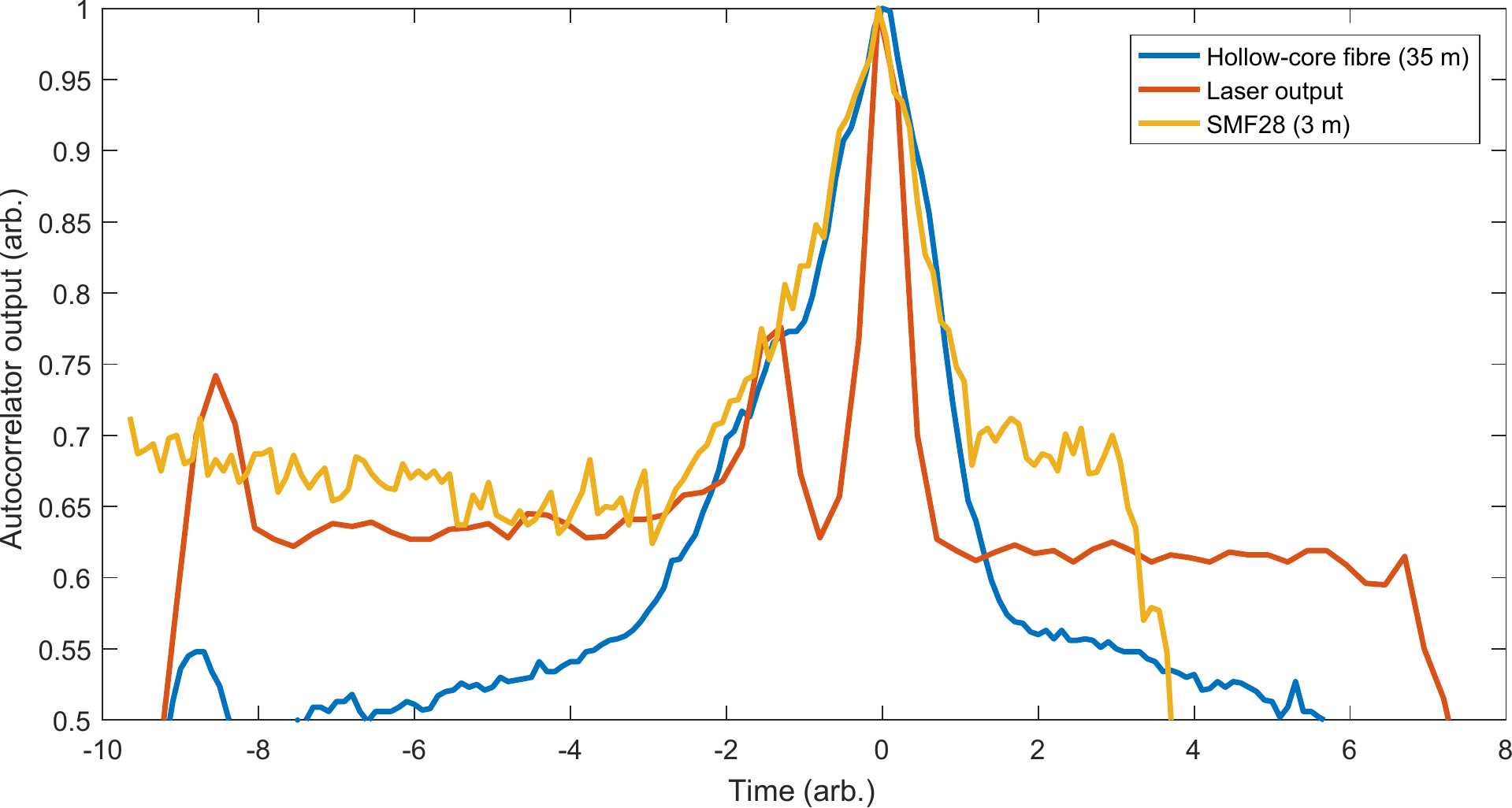}
\caption{A comparison of the dispersion of a 58~fs laser pulse due to dispersion in 35~m of hollow-core fibre and 3~m of SMF28 fibre. The comparable dispersion in the two lengths of fibre means that the hollow-core can be estimated to have approximately an order of magnitude less dispersion that the solid-core fibre (1.5~ps/nm/km).\label{fig:HCFbroadening}}
\end{figure}

For the Pritel laser used in all experiments in this paper, there was no measurable difference between the pulses before and after the HCF delay in pulse duration. This implies that for picosecond pulses the fibre is well approximated to have a no dispersive response. In order to test this further, the HCF was investigated using a Toptica FemtoFErb 1560 laser. This laser emits pulses 58~fs in length which are of order 70~nm wide in optical spectrum. The wide spectrum of the pulses will make any dispersive effects in the fibre easier to measure. For the 58~fs pulses we were able to resolve the effects of dispersion caused by the HCF. These effects are demonstrated in Figure~\ref{fig:HCFbroadening}. When compared with conventional SMF28 fibre, the dispersion in 35~m of hollow-core fibre is comparable with the dispersion in 3~m of SMF28 (which has a quoted dispersion of 18 ps/nm/km~\cite{thorlabs}). As such we can predict that the dispersion in the HCF is around an order of magnitude less than that expected from refractive index contrast fibres (1.5~ps/nm/km).

\begin{figure}[hbtp]
\centering
\includegraphics[width=0.6\linewidth]{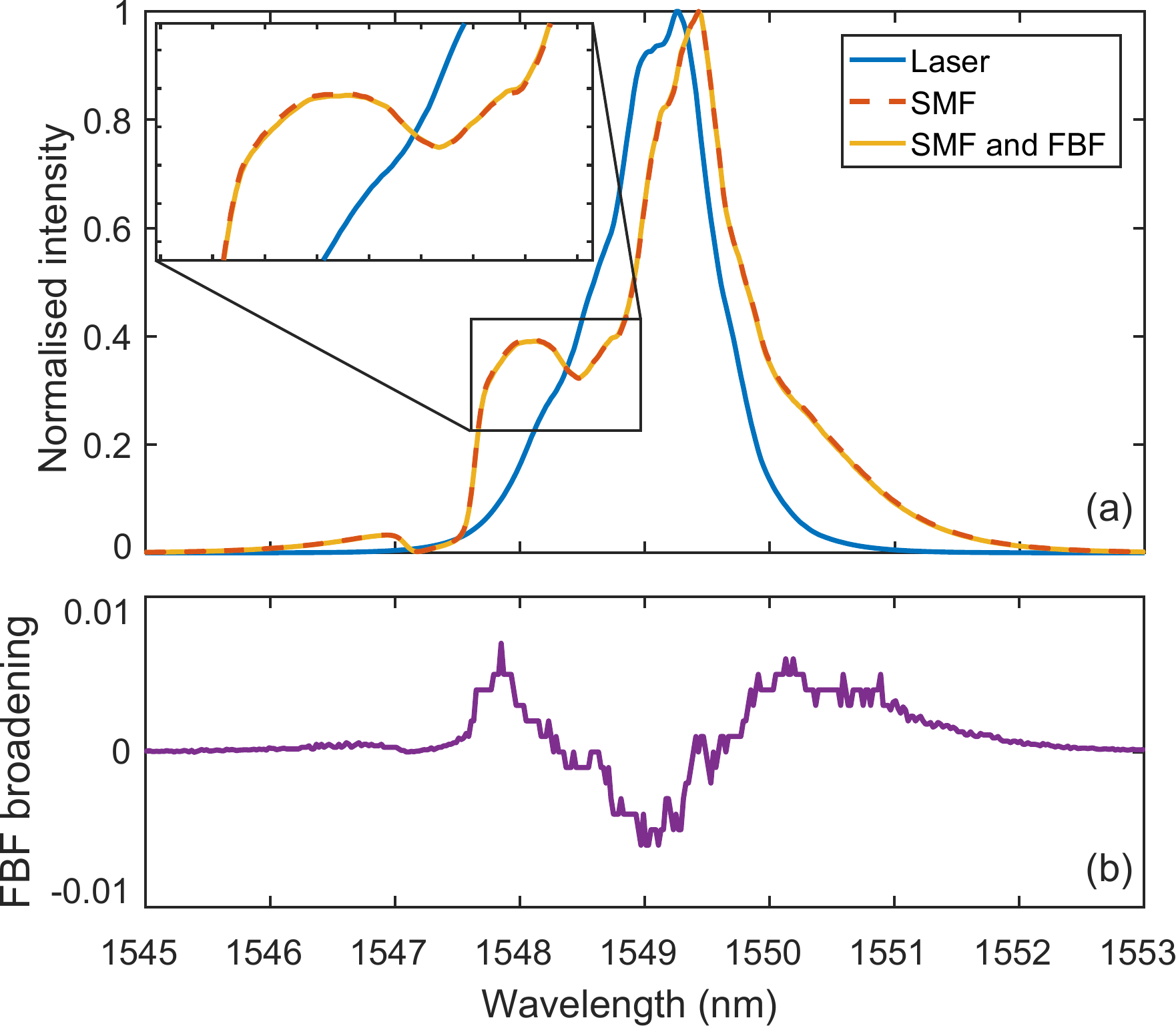}
\caption{(a) A comparison of the nonlinear spectral broadening of a 2~ps laser pulse with a peak power of 200~W due to propagation along 1~m of SMF28 fibre and 35~m of hollow-core fibre (FBF). The addition of the length of FBF adds little broadening to that induced by the single mode fibre. (b) The differences between the yellow and red lines in (a), showing a less than 1 percent change in the broadening caused by the free boundary fibre. \label{fig:spectralbroadening}}
\end{figure}

The Pritel FFL laser emits pulses with peak powers of approximately 200~W.  Figure~\ref{fig:spectralbroadening} demonstrates that for pulses of this power, the hollow core fibre fibre displays very little self-phase modulation spectral broadening and causes substantially less broadening than conventional solid core single mode fibre. Figure~\ref{fig:spectralbroadening} demonstrates  less than a 1 percent change in the broadening caused by the free boundary fibre for any particular wavelength component.

\end{document}